\documentclass[11pt,a4paper,fleqn]{article}

\usepackage{amsmath,amssymb,amsthm,enumerate}

\newcommand{\vspacebeforePopovych}{\raisebox{0ex}[2.5ex][0ex]{\null}}
\newcommand{\thetbnPopovych}{\arabic{nomer}}

\newtheorem{theorem}{Theorem}
\newtheorem{lemma}{Lemma}
\newtheorem{corollary}{Corollary}

{\theoremstyle{definition}

\newtheorem{note}{Note}
\newtheorem*{note*}{Note}
}

\begin{document}

\par\noindent {\LARGE\bf
Lie Symmetries of (1+1)-Dimensional\\
Cubic Schr\"odinger Equation with Potential\par}

{\vspace{5mm}\par\noindent {\it
Roman O. POPOVYCH~$^\dag$, Nataliya O. IVANOVA~$^\ddag$ and Homayoon ESHRAGHI~$^\star$
} \par\vspace{2mm}\par}

{\vspace{2mm}\par\noindent {\it
$^{\dag\hspace{-0.2mm},\ddag}$Institute of Mathematics of NAS of Ukraine,
3 Tereshchenkivska Str., Kyiv-4, Ukraine
} \par}
{\par\noindent {$\phantom{\dag}$~\rm E-mail: }{\it
$^\dag$rop@imath.kiev.ua, $^\ddag$ivanova@imath.kiev.ua
} \par}

{\vspace{2mm}\par\noindent {\it
$^\star$\,\,Institute for Studies in Theor. Physics and Mathematics (IPM),\\
$\phantom{^\star}$\,\,Tehran P.O.\,Box: 19395-5531, Iran,}\quad
{\rm E-mail: }{\it eshraghi@theory.ipm.ac.ir
} \par}

{\vspace{6mm}\par\noindent\hspace*{10mm}\parbox{140mm}{\small
\looseness=-2We perform the complete group classification
in the class of cubic Schr\"odinger equations of the form
$i\psi_t+\psi_{xx}+\psi^2\psi^*+V(t,x)\psi=0$,
where $V$ is an arbitrary complex-valued potential depending on $t$ and $x$.
We construct all possible inequivalent potentials
for which these equations have non-trivial Lie symmetries
using algebraic and compatibility methods simultaneously.
Our classification essentially amends earlier works on the subject.}\par\vspace{6mm}}

%\looseness=-1
Nonlinear Schr\"odinger equations (NSchEs) have a number of applications
in wave propagation in inhomogeneous media.
They arise as a model of plasma phenomena, namely,
of different processes in nonlinear and non-uniform dielectric medium
and in other branches of physics.
Schr\"odinger equations have been investigated by means of symmetry methods by
a number of authors,
see e.g.~\cite{Popovych&Ivanova:Zhdanov,Popovych&Ivanova:Nikitin,
Popovych&Ivanova:Ivanova2002,Popovych&Ivanova:Doebner1999,Popovych&Ivanova:Doebner1996,
Popovych&Ivanova:Doebner1994,Popovych&Ivanova:Gagnon88,
Popovych&Ivanova:Gagnon89c,Popovych&Ivanova:Gagnon93} and references there.
In fact, group classification for Schr\"odinger equations was first performed
by S.~Lie. More precisely, his classification~\cite{Popovych&Ivanova:Lie1881withTrans}
of all the linear equations
with two independent complex variables contains, in an implicit form,
solution of the classification problem for the linear (1+1)-dimensional
Schr\"odinger equations with arbitrary potentials.

In this paper we study a class of NSchEs of the form
\begin{equation}\label{Popovych-Ivanova:schr}
i\psi_t+\psi_{xx}+\psi^2\psi^*+V\psi=0,
\end{equation}
where the potential $V=V(t,x)$ is an arbitrary complex-valued
smooth function of the variab\-les~$t$ and~$x$.
(Here and below subscripts of functions denote differentiation
with respect to the corresponding variables.)
To find a complete set of inequivalent cases of $V$
admitting extensions of the maximal Lie invariance algebra,
we combine the classical Lie approach, studying the algebra
generated by all the possible Lie symmetry operators for equations
from class~(\ref{Popovych-Ivanova:schr}) (the adjoint representation, the inequivalent
one-dimensional subalgebras etc.)
and investigation of compatibility of classifying equations.
See~\cite{Popovych&Ivanova:Zhdanov,Popovych&Ivanova:Nikitin,
Popovych&Ivanova:Zhdanov&Lahno1999,
Popovych&Ivanova:Popovych&Yehorchenko2001a,
Popovych&Ivanova:Popovych&Ivanova2003a}
for precise formulation of group classification problems
and more details on the used methods.

Finishing excellent series of
papers~\cite{Popovych&Ivanova:Gagnon88,Popovych&Ivanova:Gagnon89c,Popovych&Ivanova:Gagnon93}
on group analysis and exact solutions of NSchEs,
in~\cite{Popovych&Ivanova:Gagnon93} L.~Gagnon and P.~Winternitz investigated essentially
more general class of variable coefficient NSchEs
than~(\ref{Popovych-Ivanova:schr}). Unfortunately, we were not able to see a direct and simple way
for deducing classifications obtained here from their results.

\begin{theorem}
Any operator
$Q=\xi^t\partial_t+\xi^x\partial_x+\eta\partial_\psi+\eta^*\partial_{\psi^*}$
from the maximal Lie invariance algebra $A^{\max}(V)$
of equation~\eqref{Popovych-Ivanova:schr} with arbitrary potential $V$
lies in the linear span of operators of the form
\begin{equation}\label{Popovych-Ivanova:gflsopscshewp}
D(\xi)=\xi\partial_t+\tfrac12\xi_tx\partial_x+\tfrac18\xi_{tt}x^2M-\tfrac12\xi_tI,\qquad
G(\chi)=\chi\partial_x+\tfrac12\chi_txM,\qquad
\lambda M.
\end{equation}
Here
$\chi=\chi(t),$ $\xi=\xi(t)$ and $\lambda=\lambda(t)$
are arbitrary smooth functions of $t,$
$M=i(\psi\partial_{\psi}-\psi^*\partial_{\psi^*})$,
$I=\psi\partial_{\psi}+\psi^*\partial_{\psi^*}.$
Moreover, the coefficients of $Q$  should satisfy the classifying condition
\begin{equation}\label{Popovych-Ivanova:classcondforcshewp}
i\eta_{\psi t}+\eta_{\psi xx}+\xi^tV_t+\xi^xV_x+\xi^t_tV=0.
\end{equation}
\end{theorem}

\begin{note}
The linear span of operators of the form~(\ref{Popovych-Ivanova:gflsopscshewp})
is an (infinite-dimensional) Lie algebra~$A^{\cup}$ under the usual Lie bracket
of vector fields. Since for any $Q\in A^{\cup}$ where $(\xi^t,\xi^x)\ne(0,0)$
we can find $V$ satisfying condition~(\ref{Popovych-Ivanova:classcondforcshewp}) then
$A^{\cup}=\langle\,\bigcup_V A^{\max}(V)\,\rangle.$
The non-zero commutation relations between the basis elements of~$A^{\cup}$
are the following ones:
\begin{gather*}
\big[D\big(\xi^1\big),D\big(\xi^2\big)\big]=D\big(\xi^1\xi^2_t-\xi^2\xi^1_t\big),\qquad
[D(\xi),G(\chi)]=G\big(\xi\chi_t-\tfrac12\,\xi_t\chi\big),\\[1ex]
[D(\xi),\lambda M]=\xi\lambda_t M,\qquad
\big[G\big(\chi^1\big),G\big(\chi^1\big)\big]=\tfrac12\big(\chi^1\chi^2_t-\chi^2\chi^1_t\big)M.
\end{gather*}
%We use the notation $\mathop{\rm Ad}\nolimits A^{\cup}$ for
%the group acting on $A^{\cup},$
%which is generated by all one-parameter groups corresponding to
%the adjoint representations of operators of~$A^{\cup}$ into~$A^{\cup}$
%and the transformation~$G(\tilde\chi)=G(-\chi)$ included additionally
%(other operators do not change).
We use the notation $\mathop{\rm Aut}\nolimits (A^{\cup})$ for
the automorphism group acting on $A^{\cup},$
which is generated by all the one-parameter groups corresponding to
the adjoint representations of operators of~$A^{\cup}$ into~$A^{\cup}$
and two discrete transformations $\mathop{\rm Ad}\nolimits I_x$
and $\mathop{\rm Ad}\nolimits I_t$ included additionally.
The actions of $\mathop{\rm Ad}\nolimits I_x$ and $\mathop{\rm Ad}\nolimits I_t$
on the basis elements of $A^{\cup}$ are defined by the formulas
$\mathop{\rm Ad}\nolimits I_x\ G(\chi)=G(-\chi)$
(the other basis operators do not change) and
$\mathop{\rm Ad}\nolimits I_t\ D(\xi)=D(\tilde\xi)$,
$\mathop{\rm Ad}\nolimits I_t\ G(\chi)=G(\tilde\chi),$
$\mathop{\rm Ad}\nolimits I_t\ \lambda M=\tilde\lambda M$,
where $\tilde\xi(t)=-\xi(-t)$, $\tilde\chi(t)=\chi(-t)$ and $\tilde\lambda(t)=-\lambda(-t)$.
\end{note}

\begin{theorem}
The Lie algebra of the kernel of maximal Lie invariance groups of equations
from class~\eqref{Popovych-Ivanova:schr} is $A^{\mathrm {ker}}=\langle M \rangle.$
\end{theorem}

\begin{theorem}
The Lie algebra~$A^{\mathop{\rm \, equiv}}$ of the equivalence group~$G^{\mathop{\rm \, equiv}}$ of
the class~\eqref{Popovych-Ivanova:schr}
is generated by the operators
\begin{gather*}
D'(\xi)=D(\xi)+\tfrac18\xi_{ttt}x^2(\partial_V+\partial_{V^*})
+\tfrac i2\xi_{tt}(\partial_V-\partial_{V^*})-
\xi_t(V\partial_V+{V^*}\partial_{V^*}),\\[1ex]
G'(\chi)=G(\chi)+\tfrac12\chi_{tt}x(\partial_V+\partial_{V^*}),\qquad
M'(\lambda)=\lambda M+\lambda_t(\partial_V+\partial_{V^*}).
\end{gather*}
Therefore, $A^{\mathop{\rm \, equiv}}\simeq A^{\cup},$
and the isomorphism is determined by means of prolongation of operators
from $A^{\cup}$ to the space~$(V,V^*).$
\end{theorem}

\begin{theorem}
The equivalence group~$G^{\mathop{\rm \, equiv}}$ of the class~\eqref{Popovych-Ivanova:schr} is generated by
the family of continuous transformations
\begin{equation}\label{Popovych-Ivanova:eqtranspcshe}
\arraycolsep=0ex\begin{array}{l}\displaystyle
\tilde t=T, \quad
\tilde x=x\varepsilon \sqrt{T_t}+X,
\quad
\tilde \psi=\psi\dfrac1{\sqrt{T_t}}
\exp\left(\dfrac i8\dfrac{T_{tt}}{T_t}\,x^2+
\dfrac i2\dfrac{X_{t}}{\sqrt{T_t}}\,x +i\Psi \right),
\\[2ex]
\tilde V=\dfrac1{T_t}\left(V+\dfrac
18\left(\dfrac{T_{tt}}{T_t}\right)_{\!\!t}x^2
+\dfrac12\left(\dfrac{X_{t}}{\sqrt{T_t}}\right)_{\!\!t}x
+\dfrac i4\dfrac{T_{tt}}{T_t}
-\left(\dfrac 14\dfrac{T_{tt}}{T_t}\,x+
\dfrac 12\dfrac{X_{t}}{\sqrt{T_t}}\right)^2
+\Psi_t\right),
\end{array}\end{equation}
and two discrete transformations:
the space reflection $I_x$
($\tilde t=t,$ $\tilde x=-x,$ $\tilde \psi=\psi,$ $\tilde V=V$)
and the Wigner time reflection $I_t$
($\tilde t=-t,$ $\tilde x=x,$ $\tilde \psi=\psi^*,$ $\tilde V=V^*$).
Here $T$, $X$ and $\Psi$ are arbitrary smooth functions of $t,$ $T_t>0.$
\end{theorem}

\begin{corollary}\label{Popovych-Ivanova:criteria.equiv.cshewp}
{\rm 1.} $G^{\mathop{\rm \, equiv}}\simeq\mathop{\rm Aut}\nolimits A^{\cup}.$
{\rm 2.} Let $A^1$ and $A^2$ be the maximal Lie invariance algebras of
equations from class~\eqref{Popovych-Ivanova:schr} for some potentials,
and ${\cal V}^i=\{\,V\,|\,A^{\max}(V)=A^i\},$ $i=1,2.$
Then ${\cal V}^1\sim {\cal V}^2\!\!\mod\!G^{\mathop{\rm \, equiv}}$
iff
$\,A^1\sim A^2\!\!\mod\!\mathop{\rm Aut}\nolimits A^{\cup}.$
\end{corollary}

\begin{lemma}\label{Popovych-Ivanova:subalgsAcup}
A complete list of $\mathop{\rm Aut}\nolimits A^{\cup}$-inequivalent
one-dimensional subalgebras of~$A^{\cup}$ is exhausted by the algebras
$\langle\partial_t\rangle,$ $\langle\partial_x\rangle,$
$\langle tM\rangle,$ $\langle M\rangle.$
\end{lemma}

\begin{proof}
Consider any operator $Q\in A^{\cup},$ i.e.\ $Q=D(\xi)+G(\chi)+\lambda M.$
Depending on the values of~$\xi,$ $\chi$ and~$\lambda$ it is equivalent
under~$\mathop{\rm Aut}\nolimits A^{\cup}$ and multiplication by a number
to one from the following operators:
$D(1)$ if $\xi\ne0;$
$G(1)$ if $\xi=0$ and $\chi\ne0;$
$tM$ if $\xi=\chi=0,$ $\lambda_t\ne0;$
$M$ if $\xi=\chi=\lambda_t=0.$
\end{proof}

\begin{corollary}\label{Popovych-Ivanova:lemma.vtvx0}
If $\,A^{\max}(V)\ne A^{\ker}\,$ then $\,V_tV_x=0\!\!\mod G^{\mathop{\rm \, equiv}}$.
\end{corollary}

\begin{proof}
Under the corollary assumption there exists
an operator $Q=D(\xi)+G(\chi)+\lambda M\in A^{\max}(V)$
which do not belong to $\langle M\rangle.$
Condition~(\ref{Popovych-Ivanova:classcondforcshewp}) implies $(\xi,\chi)\ne(0,0).$
Therefore, in force of Lemma~\ref{Popovych-Ivanova:subalgsAcup}
$\langle Q\rangle\sim \langle\partial_t\rangle$ or
$\langle\partial_x\rangle\!\!\mod\!\mathop{\rm Aut}\nolimits A^{\cup},$
i.e.\ $V_tV_x=0\!\!\mod G^{\mathop{\rm \, equiv}}$.
\end{proof}

\begin{theorem}\label{Popovych-Ivanova:theorem.gc.pcshe}
A complete set of inequivalent cases of $\,V$
admitting extensions of the maximal Lie invariance algebra
of equations \eqref{Popovych-Ivanova:schr} is exhausted by the potentials given in Table~1.
\end{theorem}

\newcounter{tbnPopovych} \setcounter{tbnPopovych}{0}
{\begin{center}
{\bf Table 1.} Results of classification.
Here $W(t),\nu,\alpha,\beta\in\mathbb{R},$ $(\alpha,\beta)\ne(0,0).$
\\[1.5ex] \footnotesize
\setcounter{tbnPopovych}{0}
\renewcommand{\arraystretch}{1.4}
\begin{tabular}{|r|c|c|l|}
\hline\vspacebeforePopovych
N &$V$ & Conditions $\!\!\mod G^{\mathop{\rm \, equiv}}$ &\hfill {Basis of $A^{\max}$\hfill} \\
\hline\vspacebeforePopovych
\thetbnPopovych & $V(t,x)$ & & $M\;$
%\hline
\\[-2.3ex]\multicolumn{4}{|c|}{\hspace*{-1.8ex}\dotfill\hspace*{-2ex}}\\[-1.8ex]  %\dotline
\raisebox{0ex}[3.3ex][0ex]{\null}%\vspacebefore
\refstepcounter{tbnPopovych}\thetbnPopovych\label{Popovych-Ivanova:pcsheV1}& $iW(t)$ & & $M,\;$ $\partial_x,\;$
$G(t)$\\[0.1ex]
\refstepcounter{tbnPopovych}\thetbnPopovych\label{Popovych-Ivanova:pcsheV2}&
$\dfrac i2\dfrac{t+\nu}{t^2+1}$ & $\nu\ge 0$
& $M,\;$ $\partial_x,\;$ $G(t),\;$ $D(t^2+1)$\\
\refstepcounter{tbnPopovych}\thetbnPopovych\label{Popovych-Ivanova:pcsheV3}&
$i\nu t^{-1}\!,\,\; \nu\ne 0,\frac12$  & $\nu\ge \frac14$
& $M,\;$ $\partial_x,\;$ $G(t),\;$ $D(t)$\\
\refstepcounter{tbnPopovych}\thetbnPopovych\label{Popovych-Ivanova:pcsheV4}& $i$ &
& $M,\;$ $\partial_x,\;$ $G(t),\;$ $\partial_t$ \\
\refstepcounter{tbnPopovych}\thetbnPopovych\label{Popovych-Ivanova:pcsheV5}& 0 &
& $M,\;$ $\partial_x,\;$ $G(t),\;$ $\partial_t,\;$ $D(t)$\\
\refstepcounter{tbnPopovych}\thetbnPopovych\label{Popovych-Ivanova:pcsheV6}& $V(x)$ &
& $M,\;$ $\partial_t$\\
\refstepcounter{tbnPopovych}\label{Popovych-Ivanova:pcsheV7}\thetbnPopovych&
$(\alpha+i\beta)x^{-2}$ & $\beta\ge 0$
& $M,\;$ $\partial_t,\;$ $D(t)$\\
\hline
\end{tabular}
\end{center}}

If we use Corollary~\ref{Popovych-Ivanova:lemma.vtvx0}, then to prove
Theorem~\ref{Popovych-Ivanova:theorem.gc.pcshe} it
is sufficient to study two cases: $V_x=0$ and $V_t=0$.
In fact, below we obtain the complete results of group classifications
for both special cases and then unite them for the general case under
consideration.

\begin{lemma} Let $V_x=0,$ i.e.\ $V=V(t).$
\begin{enumerate}

\vspace{-0.5ex}

\item[{\rm 1.}]
$A^{\mathrm {ker}}_{V_x^{\rule{0mm}{1.7mm}}=0} = \langle M,G(1),G(t)\rangle$.
$A^{\mathop{\rm \, equiv}}_{V_x^{\rule{0mm}{1.5mm}}=0} =
\langle M'(\lambda)\;
\forall\lambda\!=\!\lambda(t),G'(1),G'(t),D'(1),D'(t),D'(t^2)\rangle$.
$G^{\mathop{\rm \, equiv}}_{V_x^{\rule{0mm}{1.5mm}}=0}$
is generated by $I_t$, $I_x$ and the transformations of form~\eqref{Popovych-Ivanova:eqtranspcshe} where
$X=c_1t+c_0,$ $T=(a_1t+a_0)/(b_1t+b_0),$ \vspace{0.3ex}
$\Psi$ is an arbitrary smooth function of $t$.
$a_i,$ $b_i$ and $c_i$ are arbitrary constants such that $\,a_1b_0-b_1a_0>0.$

\vspace{-0.5ex}

\item[{\rm 2.}]
$V\sim iW\!\!\mod G^{\mathop{\rm \, equiv}}_{V_x^{\rule{0mm}{1.5mm}}=0}$ where
$W=\mathrm{Im}V.$
\vspace{0.4ex}
$\:A^{\max}(iW)\subset
A^{\cup}_{\{iW\}^{\rule{0mm}{1.5mm}}}\!=A^{\mathrm
{ker}}_{V_x^{\rule{0mm}{1.5mm}}=0}
\:\mbox{$\supset$\hspace{-1.9ex}$+$}\:
\langle D(1), D(t), D(t^2)\rangle.$
$A^{\mathrm {ker}}_{\{iW\}^{\rule{0mm}{1.5mm}}}\! =
A^{\mathrm {ker}}_{V_x^{\rule{0mm}{1.5mm}}=0}.$
$A^{\cup}_{\{iW\}^{\rule{0mm}{1.5mm}}}\! =
\langle\,\bigcup_W A^{\max}(iW)\,\rangle.$
\vspace{0.4ex}
$A^{\mathop{\rm \, equiv}}_{\{iW\}^{\rule{0mm}{1.5mm}}}\! =
\langle M,G'(1),G'(t),D'(1),D'(t),D'(t^2)\rangle$.
$\left.G^{\mathop{\rm \, equiv}}_{\{iW\}^{\rule{0mm}{1.5mm}}}\!=
G^{\mathop{\rm \, equiv}}_{V_x^{\rule{0mm}{1.5mm}}=0}\right|_{\Psi=\mathrm{const}}$.
$A^{\cup}_{\{iW\}^{\rule{0mm}{1.5mm}}}\!\simeq
A^{\mathop{\rm \, equiv}}_{\{iW\}^{\rule{0mm}{1.5mm}}}=
\mathop{\mathrm{pr}}_{(V,V^*)}A^{\cup}_{\{iW\}^{\rule{0mm}{1.5mm}}}.$

\item[{\rm 3.}]
$S=\langle D(1), D(t), D(t^2)\rangle\simeq sl(2,\mathbb{R}).$
The complete list of
$\mathop{\rm Aut}\nolimits A^{\cup}_{\{iW\}^{\rule{0mm}{1.5mm}}}\!$-inequivalent
proper subalgebras of $S$ is exhausted by
$\langle D(1)\rangle,$ $\langle D(t)\rangle,$ $\langle D(t^2+1)\rangle,$
$\langle D(1), D(t)\rangle.$

\item[{\rm 4.}]
Let $A^1$ and $A^2$ be the maximal Lie invariance algebras of
equations from class~(\ref{Popovych-Ivanova:schr}) for some potentials from $\{iW(t)\}$,
and ${\cal W}^i=\{\,W(t)\,|\,A^{\max}(iW)=A^i\},$ $i=1,2.$
Then ${\cal W}^1\sim {\cal W}^2\!\!\mod\!G^{\mathop{\rm \, equiv}}_{\{iW\}^{\rule{0mm}{1.5mm}}}$
iff
$\,A^1\bigcap S\sim A^2\bigcap S\!\!\mod\!\mathop{\rm Aut}\nolimits S.$

\item[{\rm 5.}]
If $A^{\max}_{\{iW\}^{\rule{0mm}{1.5mm}}}\! \ne
A^{\mathrm {ker}}_{V_x^{\rule{0mm}{1.5mm}}=0}$ the potential $iW(t)$
is $G^{\mathop{\rm \, equiv}}_{\{iW\}^{\rule{0mm}{1.5mm}}}\!$-equivalent
to one from Cases~2--5 of Table~1.

\end{enumerate}
\end{lemma}

\begin{note}
For any $W$ $A^{\max}(iW)\not\supset S$ (otherwise,
condition~(\ref{Popovych-Ivanova:classcondforcshewp}) would imply an incompatible system for $W$).
If $W\!=\!\mathrm{const}$ $\,W\!\in\!\{0,1\}\!\!\mod
G^{\mathop{\rm \, equiv}}_{\{iW\}^{\rule{0mm}{1.5mm}}}$
(Cases~5 and~4 correspondingly).
Cases~$2_\nu$ and~$2_{\tilde\nu}$
($3_\nu$ and~$3_{\tilde\nu}$ where $\nu,\tilde\nu\ge\frac14$)
are $G^{\mathop{\rm \, equiv}}$-inequivalent if $\nu\ne\tilde\nu.$
Since $D(t^2+1)$ cannot be contained in any two-dimensional subalgebra of $S$
it is not possible to extend $A^{\max}$ in Case~2.
There are two possibilities for extension of $A^{\max}(i\nu t^{-1})$ , namely with either
$D(1)$ (for $\nu=0$, Case~5) or $D(t^2)$ (for $\nu=\frac12$ that is equivalent to
Case~5 with respect to $G^{\mathop{\rm \, equiv}}_{\{iW\}^{\rule{0mm}{1.5mm}}}$).
\end{note}

\begin{lemma}\label{Popovych-Ivanova:classification.cshewp.Vt.0} Let $V_t=0,$ i.e. $V=V(x).$
\begin{enumerate}

\vspace{-0.5ex}

\item[{\rm 1.}]
$A^{\mathrm {ker}}_{V_t^{\rule{0mm}{1.7mm}}=0} = \langle M,D(1)\rangle$.
$A^{\mathop{\rm \, equiv}}_{V_t^{\rule{0mm}{1.5mm}}=0} =
\langle M'(1), M'(t),G'(1),D'(1),D'(t)\rangle$.
$G^{\mathop{\rm \, equiv}}_{V_t^{\rule{0mm}{1.5mm}}=0}$
consists of $I_t$, $I_x$ and the transformations of form~\eqref{Popovych-Ivanova:eqtranspcshe} where
$T_{tt}=X_t=\Psi_{tt}=0.$

\vspace{-0.5ex}

\item[{\rm 2.}]
If $A^{\max}(V) \ne
A^{\mathrm {ker}}_{V_t^{\rule{0mm}{1.5mm}}=0}$ the potential $V(x)$
is $G^{\mathop{\rm \, equiv}}_{V_t^{\rule{0mm}{1.5mm}}=0}\!$-equivalent
to one from cases of Table~2.

\end{enumerate}
\end{lemma}

\vspace{-1ex}

{\begin{center}
{\bf Table 2.} Results of classification for the subclass $\{V\!=\!V(x)\}$.
Here $\nu,\alpha,\beta\in\mathbb{R},$ $(\alpha,\beta)\ne(0,0).$
\\[1ex] \footnotesize
\renewcommand{\arraystretch}{1.3}

\begin{tabular}{|c|c|c|c|l|}
\hline\vspacebeforePopovych
N & N$_1\!\!$& $V$ & $\!\!\mod G^{\rm equiv}$ &\hfill {Basis of $A^{\max}$\hfill} \\
\hline\setcounter{tbnPopovych}{0}
\thetbnPopovych\vspacebeforePopovych& \ref{Popovych-Ivanova:pcsheV6} &
$V(x)$ & & $M,\;$ $\partial_t$
\\[-2.3ex]\multicolumn{5}{|c|}{\hspace*{-1.8ex}\dotfill\hspace*{-2ex}}\\[-1.8ex]  %\dotline
\raisebox{0ex}[3.3ex][0ex]{\null}%\vspacebefore
\refstepcounter{tbnPopovych}\label{Popovych-Ivanova:pcsheVx1}\thetbnPopovych& \ref{Popovych-Ivanova:pcsheV7} &\vspacebeforePopovych
$(\alpha+i\beta)x^{-2}$ & $\beta\ge 0$ & $M,\;$ $\partial_t,\;$ $D(t)$\\
\refstepcounter{tbnPopovych}\label{Popovych-Ivanova:pcsheVx2}\thetbnPopovych& \ref{Popovych-Ivanova:pcsheV7} &
$x^2+i+(\alpha+i\beta)x^{-2}$ & & $M,\;$ $\partial_t,\;$ $D(e^{4t})$\\
\refstepcounter{tbnPopovych}\label{Popovych-Ivanova:pcsheVx3}\thetbnPopovych& \ref{Popovych-Ivanova:pcsheV4} &
$i$ & & $M,\;$ $\partial_t,\;$ $\partial_x,\;$ $G(t)$ \\
\refstepcounter{tbnPopovych}\label{Popovych-Ivanova:pcsheVx4}\thetbnPopovych& \ref{Popovych-Ivanova:pcsheV4} &
$x+i\nu$ & $\nu>0$ &$M,\;$ $\partial_t,\;$ $G(1)+tM,\;$ $G(2t)+t^2M$\\
\refstepcounter{tbnPopovych}\label{Popovych-Ivanova:pcsheVx5}\thetbnPopovych& \ref{Popovych-Ivanova:pcsheV2} &
$-x^2+i\nu$ & $\nu\ge0$ & $M,\;$ $\partial_t,\;$ $G(\sin2t),\;$ $G(\cos2t)$\\
\refstepcounter{tbnPopovych}\label{Popovych-Ivanova:pcsheVx6}\thetbnPopovych& \ref{Popovych-Ivanova:pcsheV3} &
$x^2+i\nu,\;$ $\nu\ne\pm 1$ & $\nu\ge0$ &$M,\;$ $\partial_t,\;$ $G(e^{2t}),\;$ $G(e^{-2t})$\\
\refstepcounter{tbnPopovych}\label{Popovych-Ivanova:pcsheVx7}\thetbnPopovych& \ref{Popovych-Ivanova:pcsheV5} &
0 & & $M,\;$ $\partial_t,\;$ $\partial_x,\;$ $G(t),\;$ $D(t)$\\
\refstepcounter{tbnPopovych}\label{Popovych-Ivanova:pcsheVx8}\thetbnPopovych& \ref{Popovych-Ivanova:pcsheV5} &
$x$ &  &$M,\;$ $\partial_t,\;$ $G(1)+tM,\;$ $G(2t)+t^2M,\;$ $D(2t)+G(3t^2)+t^3M$\\
\refstepcounter{tbnPopovych}\label{Popovych-Ivanova:pcsheVx9}\thetbnPopovych& \ref{Popovych-Ivanova:pcsheV5} &
$x^2+i$ &  &$M,\;$ $\partial_t,\;$ $G(e^{2t}),\;$ $G(e^{-2t}),\;$ $D(e^{4 t})$\\
\hline
\end{tabular}
\end{center}}

\vspace{0.5ex}

\noindent {\bf Proof.} 2. Let $V=V(x)$ and
$A^{\max}(V)\ne A^{\mathrm {ker}}_{V_t=0}.$
Consider an arbitrary operator $Q=D(\xi)+G(\chi)+\lambda M\in A^{\max}(V).$
Under Lemma's assumption, the condition~(\ref{Popovych-Ivanova:classcondforcshewp}) implies
a set of equations on $V$ of the general form
\[
(ax+b)V_x+2aV=c_2x^2+c_1x+\tilde c_0+ic_0, \qquad \textrm{where}\qquad
a,b,c_2,c_1,\tilde c_0,c_0=\mathrm{const}\in\mathbb{R}.
\]
The exact number $k$ of such equations with the linear independent sets of
coefficients
can be equal to either 1 or 2.
(The value $k=0$ corresponds to the general case $V_t=0$ without
any extensions of $A^{\max}.$)

For $k=1$ $(a,b)\ne(0,0)$ and there exist two possibilities $a=0$ and $a\ne 0.$
If $a=0$ without loss of generality we can put $b=1$.
Condition~(\ref{Popovych-Ivanova:classcondforcshewp}) results in $\xi_t=0,$ $c_2=c_0=0,$ i.e.
$V_x=c_1x+\tilde c_0,$ and then $k=2$ that is impossible.

Therefore, $a\ne 0$ and we can put $a=1.$
$\tilde c_0,b=0\!\!\mod G^{\mathop{\rm \, equiv}}_{V_t=0}\!.\,$
Condition~(\ref{Popovych-Ivanova:classcondforcshewp}) results in
$\chi=0$ (then $c_1=0$), $\lambda_t=0,$ $\xi_{tt}=2c_0\xi_t$ and $c_2=c_0^2.$
Depending on $c_0=0$ and $c_0\ne 0,$ we obtain Cases~\ref{Popovych-Ivanova:pcsheVx1}
and~\ref{Popovych-Ivanova:pcsheVx2}
(of Table~2) correspondingly.

The condition~$k=2$ involves $V=d_2x^2+d_1x+\tilde d_0+id_0.\,$
$\tilde d_0=0\!\!\mod G^{\mathop{\rm \, equiv}}_{V_t=0}\!.\,$
Considering different possibilities for values of the constants $d_2,$ $d_1$ and
$d_0,$
we obtain Cases~\ref{Popovych-Ivanova:pcsheVx3}--\ref{Popovych-Ivanova:pcsheVx9}:
\begin{gather*}
d_2=d_1=d_0=0\;\to\;\ref{Popovych-Ivanova:pcsheVx7}; \qquad
d_2=d_1=0,\;d_0\ne 0\;\to\;\ref{Popovych-Ivanova:pcsheVx3}; \\
d_2=d_0=0,\;d_1\ne 0\;\to\;\ref{Popovych-Ivanova:pcsheVx8}; \qquad
d_2=0,\;d_0,d_1\ne 0\;\to\;\ref{Popovych-Ivanova:pcsheVx4}; \\
d_2<0\;\to\;\ref{Popovych-Ivanova:pcsheVx5}; \qquad d_2>0,\;d_2\ne d_0^2\;\to\;\ref{Popovych-Ivanova:pcsheVx6};
\qquad d_2>0,\;d_2=d_0^2\;\to\;\ref{Popovych-Ivanova:pcsheVx9}.
\end{gather*}

\begin{note}
To prove Theorem~\ref{Popovych-Ivanova:theorem.gc.pcshe}, it is sufficient to consider
only the case $k=1,$ $a\ne0$ in Lemma~\ref{Popovych-Ivanova:classification.cshewp.Vt.0}
since the other cases of extensions
of~$A^{\max}(V)$ with $V=V(x)$ admit operators of the form $G(\chi)+\lambda M$
($\chi\ne 0$) and, therefore (by Corollary~\ref{Popovych-Ivanova:criteria.equiv.cshewp}),
are equivalent to Cases~\ref{Popovych-Ivanova:pcsheV1}--\ref{Popovych-Ivanova:pcsheV5} of Table~1.
\end{note}

\begin{note}
The number $N_1$ for each line of Table~2 is equal to the number of
the same or equivalent case in Table~1. The corresponding equivalence
transformations have the form~(\ref{Popovych-Ivanova:eqtranspcshe})
where the functions $T,$ $X$ and $\Psi$ are as follows:
\begin{gather*}
\ref{Popovych-Ivanova:pcsheVx2}\to\ref{Popovych-Ivanova:pcsheV7},\;  \ref{Popovych-Ivanova:pcsheVx9}\to\ref{Popovych-Ivanova:pcsheV5}{:}\quad
T=-e^{-4t},\;X=\Psi=0;
\\
\ref{Popovych-Ivanova:pcsheVx6}\to\ref{Popovych-Ivanova:pcsheV3} (\tilde\nu=\tfrac{1-\nu}4){:}\quad T=-e^{-4
t},\;X=\Psi=0;
\qquad
\ref{Popovych-Ivanova:pcsheVx5}\to\ref{Popovych-Ivanova:pcsheV2} (\tilde\nu=\nu){:}\quad T=\tan 2t,\;X=\Psi=0;
\\
\ref{Popovych-Ivanova:pcsheVx8}\to\ref{Popovych-Ivanova:pcsheV5}{:}\quad T=t,\;X=-t^2,\;\Psi=\dfrac{t^3}3;
\qquad
\ref{Popovych-Ivanova:pcsheVx4}\to\ref{Popovych-Ivanova:pcsheV4}{:}\quad
T=|\nu|t,\;X=-\sqrt{|\nu|}\,t^2,\;\Psi=\dfrac{t^3}3.
\end{gather*}
\end{note}

\medskip

The results of the group classification obtained in this paper can be used
to construct both invariant and partially invariant exact solutions
of equations having the form~(\ref{Popovych-Ivanova:schr}).
Moreover, we plan to study conditional symmetries of~(\ref{Popovych-Ivanova:schr})
to find non-Lie exact solutions.

Another direction for our future research to develop the above results is investigation
of a more general class of $(1+n)$-dimensional NSchEs with potentials
\begin{equation}\label{Popovych-Ivanova:generalNSchEwithPotential}
i\psi_t+\Delta\psi+F(\psi,\psi^*)+V(t,\vec x)\psi=0,
\end{equation}
where the $F=F(\psi,\psi^*)$ is an arbitrary complex-valued
smooth function of the variables~$\psi$ and~$\psi^*$.
We have already described all possible inequivalent forms of the
parameter-function~$F$ (without any restriction on the dimension $n$)
for which an equation of the form~(\ref{Popovych-Ivanova:generalNSchEwithPotential})
with a~some potential $V$ has an extension of the maximal Lie invariance
algebra.
We believe that the classification method suggested in this paper
can be effectively applied to complete the group classification
in~(\ref{Popovych-Ivanova:generalNSchEwithPotential}) for the small values of $n.$

\subsection*{Acknowledgements}

The authors are grateful to
Profs.~V.~Boyko, A.~Nikitin, I.~Yehorchenko and A.~Zhalij
for useful discussions and interesting comments.
The research of NMI was partially supported by National Academy of Science of Ukraine
in the form of the grant for young scientists.
ROP is grateful to Prof.~F.~Ardalan
(School of Physics, Institute for Studies in Theoretical Physics and Mathematics, Tehran)
for hospitality and support during writing this paper.

%\LastPageEnding

\end{document}